\documentclass{ws-procs9x6}

\newcommand{\intsum}{\hspace{2mm}\int\hspace*{-5mm}\sum}

\begin{document}

\title{Modification of the Casimir Effect\\ due to a Minimal length scale}

\author{Ulrich Harbach}

\address{Institut f{\"u}r Theoretische Physik\\
Johann Wolfgang Goethe-Universit{\"a}t\\
and\\
Frankfurt Institute for Advanced Studies\\
Max-von-Laue-Str. 1\\
60438 Frankfurt am Main, Germany\\ 
E-mail: harbach@th.physik.uni-frankfurt.de}

\author{Sabine Hossenfelder}

\address{Department of Physics\\
University of Arizona\\
1118 East 4th Street\\
Tucson, AZ 85721, USA\\
E-mail: sabine@physics.arizona.edu}  

\maketitle

\abstracts{The existence of a minimal length scale, a fundamental lower limit on spacetime resolution is motivated by various theories of quantum gravity as well as string theory. Classical calculations involving both quantum theory and general relativity yield the same result. This minimal length scale is naturally of the order of the Planck length, but can be as high as $\sim {\rm TeV}^{-1}$ in models with large extra dimensions. We discuss the influence of a minimal scale on the Casimir effect on the basis of an effective model of quantum theory with minimal length.}

\section{The minimal length scale}
\subsection{Motivation}
The idea of a minimal length has a long history and was already discussed by W. Heisenberg in the 1930s, who recognised its importance in regularising UV-divergences\cite{Heisenberg:1938}. Today, theories beyond the standard model such as string theory or loop quantum gravity - as diverse as they may be - all suggest the existence of a fundamental limit to spacetime resolution of the order of the Planck length. Thus, the motivations for the existence of a minimal length scale are manifold:
\begin{itemize}
    \item In perturbative string theory\cite{Gross:1988ar,Amati:1988tn}, the feature of a fundamental minimal length scale arises from the fact that strings cannot probe distances smaller than the inverse string scale. If the energy of a string reaches this scale $M_s=\sqrt{\alpha'}$, excitations of the string can occur and increase its extension\cite{Witten:1997fz}. In particular, an examination of the spacetime picture of high-energy string scattering shows, that the extension of the string is proportional to its energy\cite{Gross:1988ar} in every order of perturbation theory. Due to this, uncertainty in position measurement can never become arbitrarily small.
    \item In loop quantum gravity, spacetime itself is quantised and thus measurements of area and volume at small scales must fall into the spectrum of the respective self-adjoint operators, which is discrete\cite{Rovelli:1994ge}.
    \item Including gravitational effects from general relativity into a classical analysis of the process of position measurement yields a minimal uncertainty\cite{Mead:1964}, i.e. a minimal length is implicitly contained in the standard model (SM) combined with general relativity.
\end{itemize}

\subsection{Large extra dimensions}
Arkani-Hamed, Dimopoulos and Dvali proposed a solution to the hierarchy problem (the hugeness of the Planck scale compared to the scale of electroweak symmetry breaking) by the introduction of $d$ additional compactified spacelike dimensions in which only the gravitons can propagate\cite{Arkani-Hamed:1998rs,Antoniadis:1998ig}. The SM particles are bound to our 4-dimensional sub-manifold, often called our 3-brane. Due to its higher dimensional character, the gravitational force at small distances then is much stronger in these models. This results in a lowering of the Planck scale to a new fundamental scale, $M_{\rm f}$, which can be as low as the TeV-range. Accordingly, in such models the minimal length scale increases to a new fundamental length scale $L_{\rm f}$.

\section{Quantum theory with minimal length}
To include effects of the minimal length, we assumethat at arbitrarily high momentum $p$ of a particle, its wavelength is bounded by some minimal length $L_{\mathrm f}$ or, equivalently, its wave-vector $k$ is bounded by a $M_{\mathrm f}=1/L_{\rm f}$\cite{Ahluwalia:2000iw}. Thus, the relation between the momentum $p$ and the wave vector $k$ is no longer linear $p=k$ but a function $k=k(p)$\footnote{Note, that this is similar to introducing an energy dependence of Planck's constant $\hbar$.}, which has to fulfil the following properties\cite{Hossenfelder:2003jz,Hossenfelder:2004up}:
\begin{enumerate}
\item[a)]  For energies much smaller than the new scale it yields the linear relation:
for $p \ll M_{\mathrm f}$ we have $p \approx k$. \label{limitsmallp}
\item[b)] It is an an uneven function (because of parity) and $k \parallel p$.
\item[c)]  The function asymptotically approaches the bound $M_{\mathrm f}$. \label{upperbound}
\end{enumerate}
The quantisation in this scenario is straightforward and follows the usual procedure. Using the well known commutation relations
\begin{eqnarray} \label{CommXK}
[\hat x_i,\hat k_j]={\mathrm i } \delta_{ij}\quad
\end{eqnarray}
and inserting the functional relation between the wave vector and the momentum then yields the modified commutator for the momentum and results in the generalized uncertainty principle ({\sc GUP})
\begin{eqnarray} \label{CommXP}
[\,\hat{x}_i,\hat{p}_j]&=& + {\rm i} \frac{\partial p_i}{\partial k_j} \quad\longrightarrow\quad \Delta p_i \Delta x_j \geq \frac{1}{2}  \Bigg| \left\langle \frac{\partial p_i}{\partial k_j}\right\rangle \Bigg| \quad,
\end{eqnarray}
which reflects the fact that it is not possible to resolve space-time distances arbitrarily well. Because $k(p)$ becomes asymptotically constant, its derivative $\partial k/ \partial p$ eventually vanishes and the uncertainty (Eq.(\ref{CommXP})) increases for high momenta. Thus, the introduction of the minimal length reproduces the limiting high energy behavior found in string theory\cite{Gross:1988ar}.

In field theory\footnote{For simplicity, we consider a massless scalar field.}, one imposes the commutation relations Eq. (\ref{CommXK}) and (\ref{CommXP}) on the field $\phi$ and its conjugate momentum $\Pi$. Its Fourier expansion leads to the annihilation and creation operators which must obey
\begin{eqnarray}
\left[\hat{a}_k,\hat{a}^\dag_{k'}\right] &=& - {\rm i}
\left[\hat{\phi}_k,\hat{\Pi}^\dag_{k'}\right] \quad ,\\
\left[\hat{a}_k,\hat{a}^\dag_{k'}\right] &=& \delta(k-k') \quad ,\\
\left[\hat{a}_p,\hat{a}^\dag_{p'}\right] &=& \Bigg| \frac{\partial k}{\partial p}
\Bigg| ~\delta(p-p')  \quad. \label{coma}
\end{eqnarray}

\section{The Casimir energy}
Zero-point fluctuations of any quantum field give rise to observable Casimir forces if boundaries are present\cite{Casimir:1948dh}. Here, we consider the case of two conducting parallel plates in a distance $a$ in direction $z$. Using the framework developed above, in the presence of a minimal length the vacuum expecation value (VEV) for the field energy density is now given by\cite{Harbach:2005yu}
\begin{eqnarray}
\langle 0 \vert\hat{H}\vert 0 \rangle &=& \langle 0 \vert\frac{1}{2} \intsum {\mathrm d}^{3} p \; \left( \hat{a}^{\dag}_p \hat{a}_p
+ \hat{a}_p \hat{a}^{\dag}_p \right) E \vert 0 \rangle\nonumber\\
 &\approx& \frac{1}{2} \intsum {\mathrm d}^{3} p \; \exp\bigg ({\displaystyle{-\frac{L_{\mathrm f}^2 \pi}{4} p^2}}\bigg ) E \quad,
\end{eqnarray}
where $E$ is the energy of a mode with momentum $p$. Here, we have used the specific relation from Ref. \refcite{Hossenfelder:2004up} for $k(p)$
\begin{eqnarray}
k_{\mu}(p) &=& e_{\mu} \int_0^{p} \exp\bigg ({\displaystyle{-\frac{L_{\mathrm f}^2 \pi}{4} p^2}}\bigg ) \label{model} \quad,
\end{eqnarray}
where $e_{\mu}$ is the unit vector in $\mu$ direction. It is easily verified that this expression fulfills the requirements (a) - (c).

To obtain the Casimir energy, the difference of the VEVs of the inside and the outside regions of the plates has to be taken:

For Minkowski space in $3+1$ dimensions without boundaries, the energy density in the present model with minimal length is finite due to the squeezed momentum space at high momenta and given by
\begin{eqnarray} \label{mink}
\varepsilon_{\mathrm{Mink}} = \langle 0 \vert \hat{H} \vert 0 \rangle = \frac{16}{\pi} \frac{M_{\rm f}}{L_{\rm f}^3} \quad.
\end{eqnarray}

The quantisation of the wavelengths between the plates in the $z$-direction yields the condition $k_l = l/a$. Since the wavelengths can no longer get arbitrarily small, the smallest wavelength possible belongs to a finite number of nodes $l_{\rm max}$. As a result, momenta come in steps $p_l = p (k_l)$ which are no longer equidistant $\Delta
p_l = p_l - p_{l-1}$. Then
\begin{eqnarray} \label{Hplates}
\varepsilon_{\mathrm{Plates}}=\pi \hspace*{-2mm} \sum_{l=-l_{\rm max}}^{l_{\rm max}} \hspace*{-2mm}
\Delta p_l \int_0^{\infty} {\mathrm d} p_{\parallel}   \; \;
e^{{\displaystyle{-\epsilon p_{\parallel}^2}}}
e^{{\displaystyle{-\epsilon p_l^2}}}  E~p_{\parallel} \quad,
\end{eqnarray}
where $p_{\parallel}^2=p_x^2 + p_y^2$ and $E^2 = p_{\parallel}^2 +p_l^2$.

\begin{figure}
\vspace{-0.8cm}
  \begin{center}
    \begin{minipage}[t]{0.63\linewidth}
      \raisebox{-4.5cm}{\includegraphics[width=\linewidth]{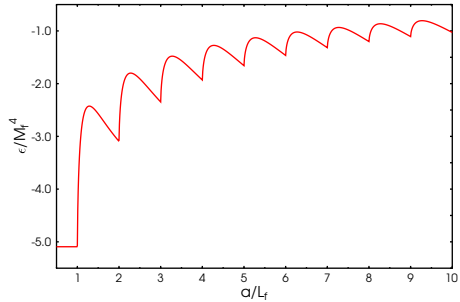}}
    \end{minipage}
    \begin{minipage}[t]{0.28\linewidth}
      \caption{The Casimir energy density between two plates of distance $a$ in units of the minimal length. \label{fig1}}
    \end{minipage}\hfill
  \end{center}
\vspace{-0.8cm}
\end{figure}

The result of our calculation is shown in Fig. \ref{fig1}. The slope of the curve changes whenever another mode fits between the plates. Although the slope (and thus the Casimir force) is singular at these points, the plot clearly shows that a finite energy is sufficient to surmount them and thus the result is physical. These singularities result from the assumption of two strictly localised plates and might be cured in a full theory by the minimal length uncertainty on the plate positions.

\section{Conclusion}
The existence of a minimal length scale is justified on various grounds. The minimal length is considerably increased in models with large extra dimensions. We presented an effective model that incorporates the minimal length into quantum theory. As an application, the Casimir energy for two parallel plates was studied. This example depicts nicely how the minimal length acts as a natural regulator for infinities in quantum field theories.

\section*{Acknowledgments}
U.H. thanks the Frankfurt Institute of Advanced Studies for financial support through a PhD scholarship, Marcus Bleicher for fruitful discussions and the provision of funds for the travel, and the organisers of the LLWI for a wonderful conference. S.H. acknowledges support by the {\sc DFG} and NSF PHY/0301998.

\bibliography{bormio}

\begin{thebibliography}{10}

\bibitem{Heisenberg:1938}
W. Heisenberg,
\newblock Ann. Phys. 32 (1938) 20.

\bibitem{Gross:1988ar}
D.J. Gross and P.F. Mende,
\newblock Nucl. Phys. B303 (1988) 407.

\bibitem{Amati:1988tn}
D. Amati, M. Ciafaloni and G. Veneziano,
\newblock Phys. Lett. B216 (1989) 41.

\bibitem{Witten:1997fz}
E. Witten,
\newblock Phys. Today 50N5 (1997) 28.

\bibitem{Rovelli:1994ge}
C. Rovelli and L. Smolin,
\newblock Nucl. Phys. B442 (1995) 593, gr-qc/9411005.

\bibitem{Mead:1964}
C.A. Mead,
\newblock Phys. Rev. 135 (1964) B849.

\bibitem{Arkani-Hamed:1998rs}
N. Arkani-Hamed, S. Dimopoulos and G.R. Dvali,
\newblock Phys. Lett. B429 (1998) 263, hep-ph/9803315.

\bibitem{Antoniadis:1998ig}
I. Antoniadis et~al.,
\newblock Phys. Lett. B436 (1998) 257, hep-ph/9804398.

\bibitem{Ahluwalia:2000iw}
D.V. Ahluwalia,
\newblock Phys. Lett. A275 (2000) 31, gr-qc/0002005.

\bibitem{Hossenfelder:2003jz}
S. Hossenfelder et~al.,
\newblock Phys. Lett. B575 (2003) 85, hep-th/0305262.

\bibitem{Hossenfelder:2004up}
S. Hossenfelder,
\newblock (2004), hep-ph/0405127.

\bibitem{Casimir:1948dh}
H.B.G. Casimir,
\newblock Kon. Ned. Akad. Wetensch. Proc. 51 (1948) 793.

\bibitem{Harbach:2005yu}
U. Harbach and S. Hossenfelder,
\newblock (2005), hep-th/0502142.

\end{thebibliography}
\bibliographystyle{h-elsevier2}

\end{document}